\pdfoutput=1
\documentclass[aps,prb,preprint,showpacs,floatfix]{revtex4}
\usepackage{dcolumn}
\usepackage{bm}
\usepackage{graphicx}
\usepackage{subfigure}
\begin{document}

\title{Atomistic simulation of frictional anisotropy on quasicrystal approximant surfaces}

\author{Zhijiang Ye}
\affiliation{Department of Mechanical Engineering, University of California Merced, 5200 N. Lake Road, Merced, CA 95343, USA}

\author{Ashlie Martini}
\affiliation{Department of Mechanical Engineering, University of California Merced, 5200 N. Lake Road, Merced, CA 95343, USA}

\author{Patricia Thiel}
\affiliation{Departments of Chemistry and Materials Science and Engineering,
Iowa State University and the Ames
Laboratory, Ames, IA}

\author{Heather H.~Lovelady}
\author{Keith McLaughlin}
\author{David A.\ Rabson}
\email[]{davidra@ewald.cas.usf.edu}
\affiliation{Department of Physics, University of South Florida, 4202 E. Fowler Ave., Tampa, FL 33617, USA}

\date{\today}

\begin{abstract}

Park {\it et al.} have reported eight times greater atomic-scale friction
in the periodic than in the quasiperiodic direction
on the two-fold face of a decagonal Al-Ni-Co quasicrystal [Science {\bf309},
1354 (2005)].
We present results of molecular-dynamics simulations intended to elucidate
mechanisms behind this giant frictional anisotropy.  Simulations of
a bare atomic-force-microscope tip on several model substrates
and under a variety of conditions failed to reproduce experimental results.
On the other hand, including the experimental passivation of the
tip with chains of hexadecane thiol, we reproduce qualitatively
the experimental anisotropy in friction, finding
evidence for entrainment of the organic chains in surface furrows parallel
to the periodic direction.

\leftline{Published: Phys.\ Rev.\ B \bf93\rm, 235438 (2016)}

\end{abstract}


\maketitle

\section{Introduction}

The peculiarly low friction between quasicrystal surfaces
and a contacting probe remains a puzzle.\cite{rabson2012review}
Most of the proposed and realized applications of quasicrystals have taken
advantage of this low friction.\cite{Patent88,fryingpans94,SymkoPatent,dubois2000new,dubois11israel}
Anomalously low friction has been reported
at length scales ranging from atomic probes to engineering pin-on-disk
experiments, in air and in vacuum, and at such low normal forces as to
avoid any surface damage, even at the atomic scale, as well as in ploughing
experiments.\cite{park08b}  Quasicrystals exhibit lower friction
than related periodic phases, called approximants, of similar chemical
composition,\cite{Mancinelli03,park08b} suggesting but not proving
that more than surface chemistry is involved.
At the same time, the lack of periodicity
in quasicrystals makes them hard, and low friction could be a consequence of
hardness.\cite{park08b,Urban93,Urban99,brunet00,dubois2000new}
A 2005 experiment by Park {\it et al.} sought to dispense
with all such surface-dependent effects by measuring the friction between a
thiol-passivated atomic force microscope (AFM) probe and a two-fold surface
of a decagonal AlNiCo quasicrystal.\cite{park2005high}  On this surface,
one direction is periodic, the other quasiperiodic.  That surface damage
was averted through the intervening thiol was demonstrated by STM imaging
before and after the friction measurements.  The experiment, in
ultra-high vacuum, found an eight-fold anisotropy
in the magnitude of sliding friction.  

Such a large surface-friction anisotropy was virtually unknown in atomic-scale
measurements; a recent review of theories of quasicrystal friction dubbed it
the ``giant frictional-anisotropy effect.''\cite{rabson2012review}
Filippov {\it et al.},
attacking the problem computationally, reported reproducing the
experimental anisotropy in a Langevin model in which mean feature spacings
differed in the periodic and quasiperiodic directions; the quasiperiodicity itself
was not relevant.\cite{filippov2010origin,filippov10E,filippovreply11}
However, three of us, using the
same methods, found that small changes in the parameters could change the
sense of the anisotropy and argued, moreover, that Filippov's parameters did
not correspond to scanning-tunneling-microscope (STM) images of the experimental surface.\cite{mclaughlin11}

Other plausible explanations include entrainment of the thiol chains
passivating the AFM tip and the difficulty of either exciting or propagating
phonon or electronic modes.
In such a complex system, it would
not be surprising if each of these mechanisms played a role; on the other
hand, the generic result of low friction in different quasicrystalline
materials under different circumstances suggests a generic mechanism tied
to quasicrystallinity.  Controversy over the frictional anisotropy on d-AlNiCo
echoes controversy over the various contributions to atomic-scale friction on
other surfaces that do not combine periodic and quasiperiodic order.\cite{Tobin93,Sokoloff95,krim95,krim98}
By testing each of the proposed mechanisms (except electronic),
molecular-dynamics (MD) simulation on d-AlNiCo may provide clues to the
nature of atomic-scale friction generally,\cite{dong2013review} and that
is the purpose of the current paper.

We report MD friction simulations on two quasicrystal approximants as well
as on a series of artificial ``Fibonaccium'' samples approaching true
quasiperiodicity.  We searched for, but did not find, evidence 
for the phonon hypothesis.
We find instead that simulation of the thiol molecules
passivating the tip is necessary to produce substantial frictional
anisotropy, and by examining surface topography, thiol configurations,
and adhesion forces, argue that entrainment in furrows is responsible
for at least some of the observed anisotropy.

\section{Methods}

Perhaps the only simple aspect of the surfaces of quasicrystals is that
they appear to reflect the underlying bulk structures without reconstruction.
\cite{McGrath10}  While these materials exhibit a high degree of long-range
positional order, as demonstrated in Fourier space by very sharp diffraction
peaks,\cite{Kycia93} they do so without periodicity; in one sense, their
unit cells are infinite.  As a consequence, only a few bulk structures
have been proposed at a level permitting comparison to experiment.\cite{takakura07}  This poses a particular challenge to the simulation of
atomic-scale friction on quasicrystal surfaces.  However, quasicrystalline
approximants contain local symmetries and structural motifs similar to those
of their quasicrystalline counterparts \cite{henley1991cell} and are more
suitable for MD simulations because of their periodicity.

In this work, we used two d-AlNiCo approximants supplied by Widom,
H1 with a unit cell of 25 atoms and T11
with 343 atoms per unit cell.\cite{Widomprivate}
These crystal structures were repeated in space to create model substrates.
The apex of an AFM tip was then introduced into the models to enable
simulations of sliding friction.  The size and sliding surface of the
approximant sample and the size, shape, and material of the tip were varied.
The tip was subjected to a normal load and slid across the substrate surface
in two orthogonal directions, corresponding to periodicity and (approximate)
quasiperiodicity on the sample surface.  Friction was then calculated as the
time average of the lateral force resisting that sliding.  In all cases,
temperature was controlled using a Langevin thermostat, and the simulations
were run using the LAMMPS simulation software.\cite{plimpton1995fast}
In what follows, references to the ``quasiperiodic'' direction are intended
in the sense of an approximant, thus referring to the larger of the two
periodicities.

\section{Adamant Tip on d-AlNiCo Approximants}

\begin{figure}
\begin{center}
%	\subfigure[]{\includegraphics[bb=0 0 800 842, trim=0 0 50 200, clip, scale=0.28]{fullmodern_rev4.pdf}}
	\subfigure[]{\includegraphics[scale=0.36]{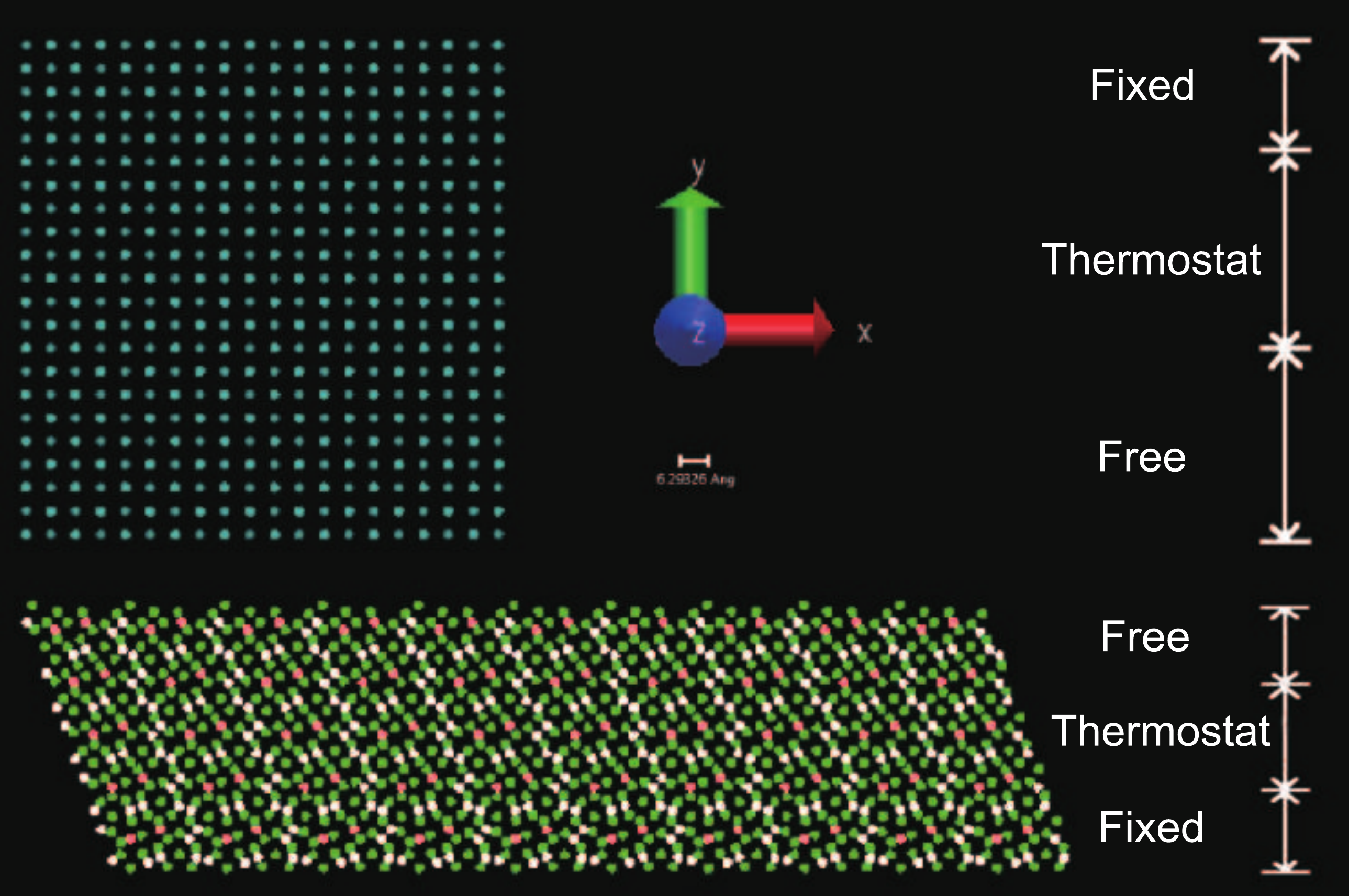}}
	\subfigure[]{\includegraphics[scale=0.22]{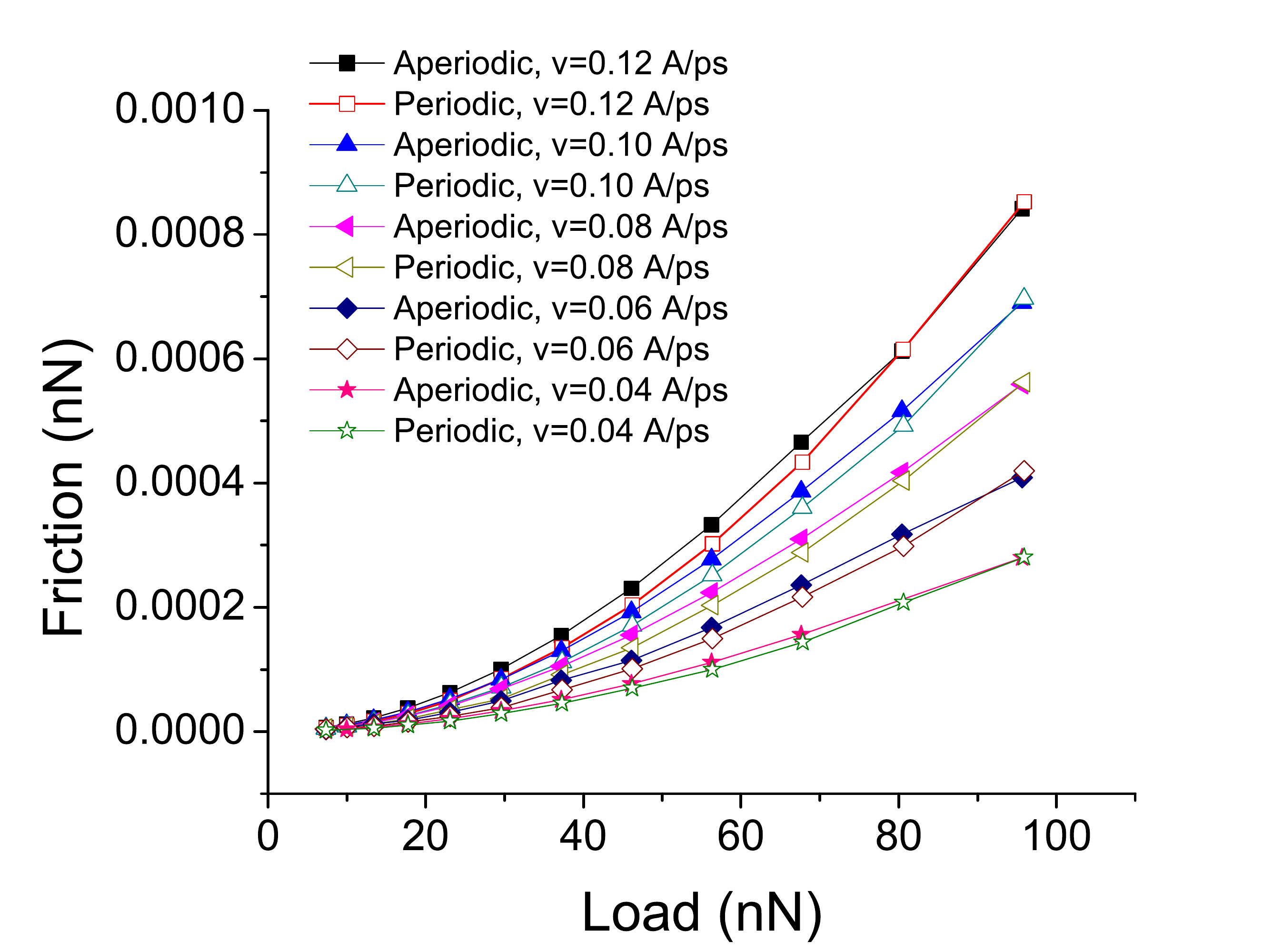}}
\end{center}
\caption{(a) Snapshot of a run on the H1 approximant showing adamant tip upper-left
and facing the 10-fold surface of the approximant.  The $x$ direction is the
(approximately) quasiperiodic direction for the 2-fold surface, $z$ the
periodic direction.  Green spheres are Al, white Ni, pink Co, and blue
adamant.
(b) Average friction force increasing with load from a
simulation of an adamant slab sliding on an H1 approximant. Anisotropy is
small and, in most cases, the trend is opposite the experimentally observed
anisotropy; the
magnitudes of the friction force and friction coefficient are also much
smaller than expected.} \label{fig:heather} \end{figure}

The first model we discuss consisted of 1,250 unit cells of the 25-atom H1
approximant (31,250 atoms) and an ``adamant'' slab tip (3,250 atoms) moving
laterally across its surface (Fig.~\ref{fig:heather}a).\cite{Harper08}
Interactions among the Al, Ni, and Co atoms of the approximant were described
by a set of pair potentials designed for this system by Widom and Moriarty,
\cite{moriarty1997first,widom2000first,widom1998first} truncated at 0.7~nm.
The adamant of the
tip was a fictitious atom of the same mass as aluminum forming a very hard
and inert FCC solid.  Adamant-adamant pair interactions were given by the
Widom-Moriarty Al-Al potential multiplied by 10, while the purely repulsive
force with any atom in the approximant was $b\exp(-r/a)$, $r$ the pair
separation, $b=100\,$eV, and $a=1\,$\AA.  The top-most layers of the tip were
treated as a rigid body and the bottom-most layers of the approximant fixed.
The center layers of the tip and approximant were subject to the thermostat
in the direction perpendicular to sliding to control the temperature of the
simulation at less than 1K.\cite{endnote}
%\footnote{An isotropic thermostat at an earlier
%stage of the investigation introduced significant noise; restricting
%the thermostat to the $z$ direction\cite{SinnottPrivate,Gosvami11} reduced the
%noise.  If the phonon hypothesis had been correct, the lateral
%thermostat could have masked the effect.}
The remaining atoms in the system were free
to evolve according to NVE dynamics (constant number, constant volume,
and constant energy).  A normal force was applied by moving
the tip downward towards the substrate after which the tip was slid across
the substrate at 5 m/s.

Fig.~\ref{fig:heather} shows results for loads between 7.3 nN and 95.7 nN
and sliding speeds between 4 and 12 m/s.  For each speed and at lower loads,
the frictional force in the quasiperiodic direction
was slightly higher than in the periodic direction.
This trend is opposite that expected based on the experiment.  Using the
higher loads, the ratios of lateral to normal force in Fig.~\ref{fig:heather},
{\it i.e.}, coefficient of friction, could be fit linearly, and there was
a small anisotropy in the expected direction with the periodic coefficient
of friction higher than quasiperiodic by about 11\%.  However, both frictional
force and frictional coefficient were orders of magnitude smaller than
those reported experimentally.

\begin{figure}\begin{center}
\includegraphics[scale=0.25]{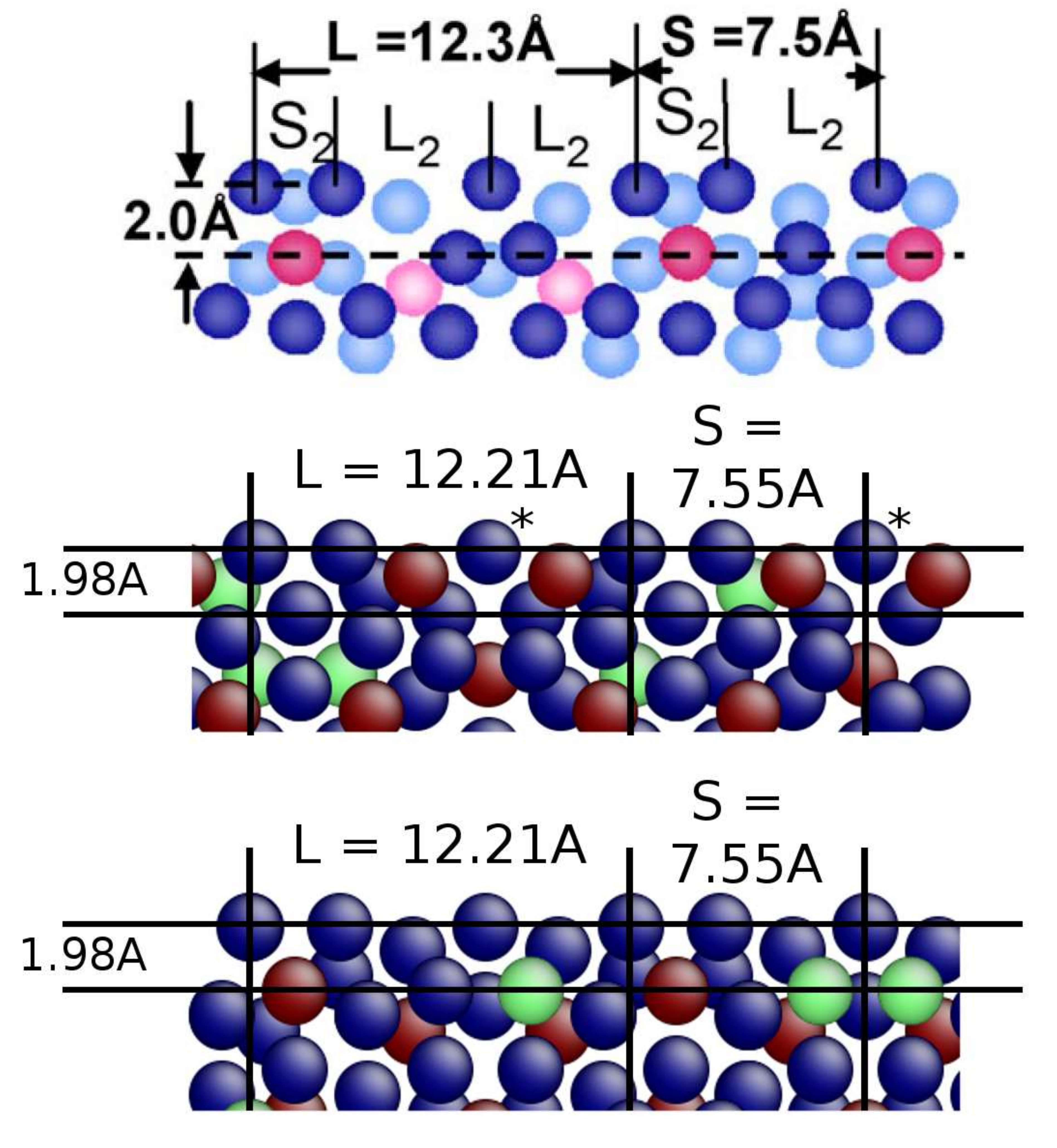}
\end{center}
\caption{(Top) Experimental surface model from [\onlinecite{Park05b}].
The blue atoms
are aluminum, pink transition metals.  The T11 surface terminations in the
middle and bottom panels both show features in common with the experimental
model.  Here, blue atoms are aluminum, red cobalt, and green nickel.  Atoms
marked (*) in the middle panel were unstable and so removed before friction
simulation.}
\label{fig:keithT11}
\end{figure}

\begin{figure} \begin{center}
      \subfigure[]{\includegraphics[scale=0.25]{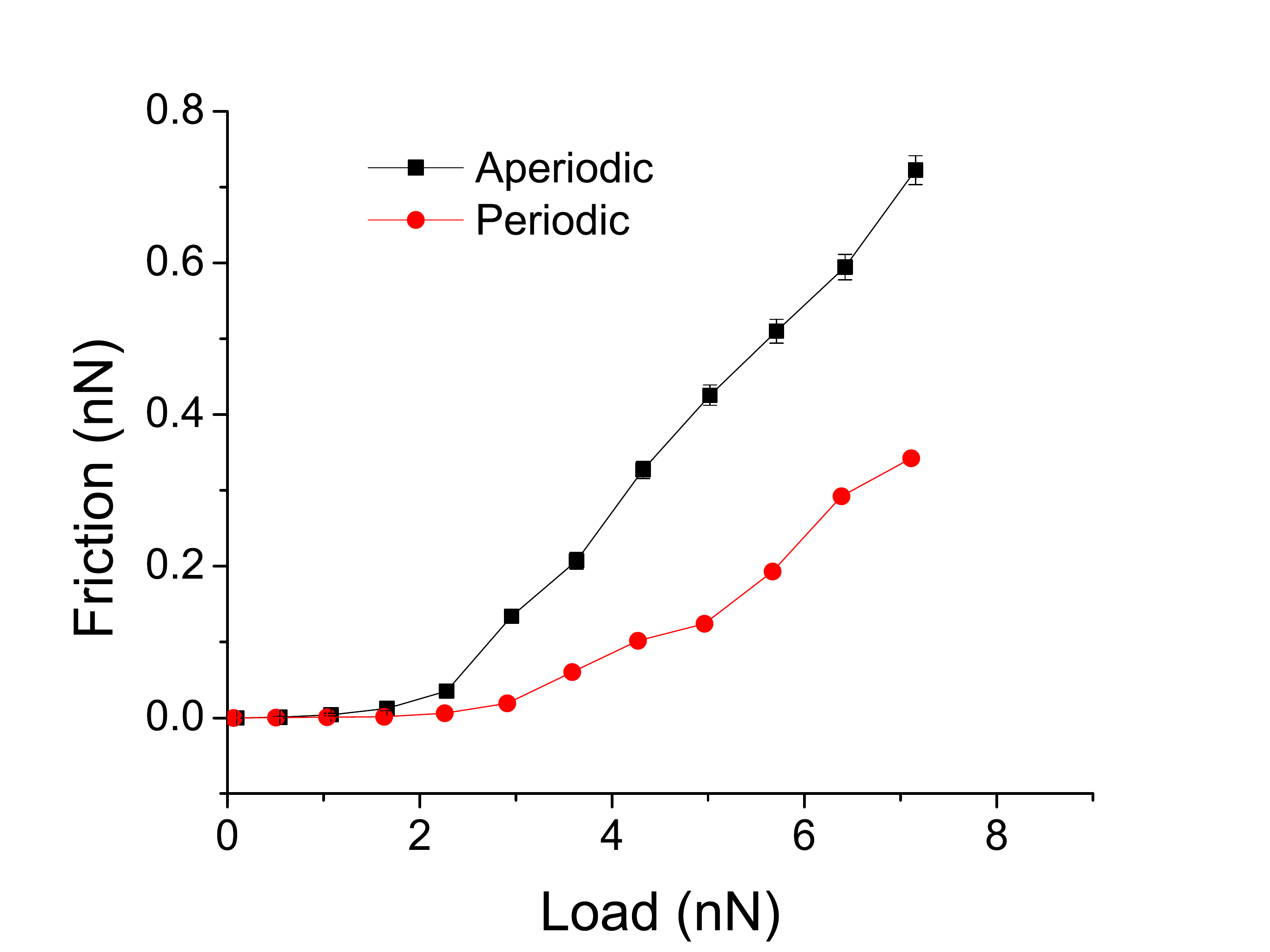}}
      \subfigure[]{\includegraphics[scale=0.25]{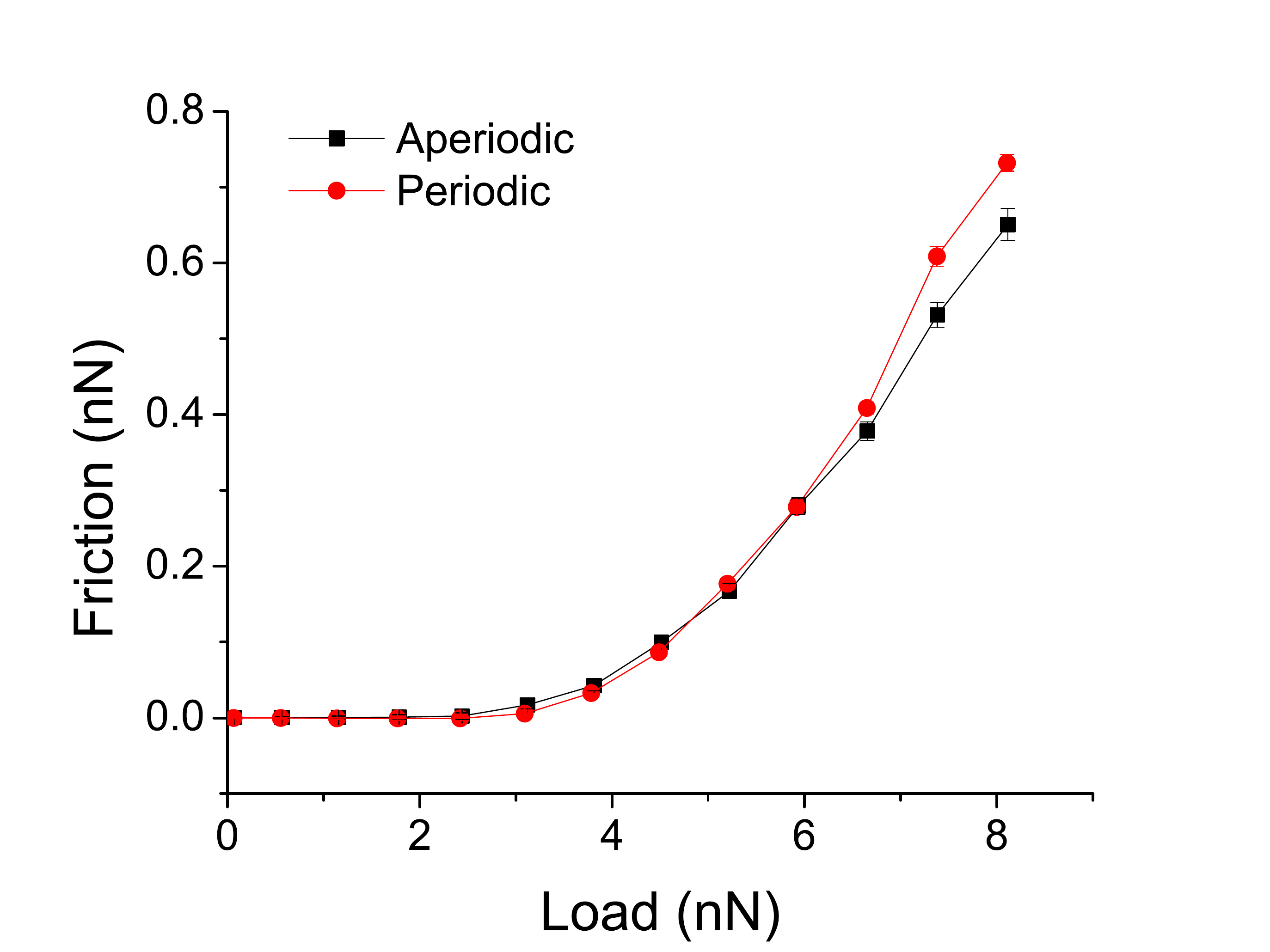}}
\end{center} \caption{Average friction force increasing with load from a
simulation of a rigid FCC aluminum block sliding on terminations T11(a) and
T11(b) of Figure~\ref{fig:keithT11}.
Although the magnitude of the friction is realistic,
the frictional anisotropy seen in experiments is not reproduced.} \label{fig:keith1}
\end{figure}

To improve the physical realism of the simulation, we replaced the H1
approximant with the larger-unit-cell T11 approximant.  The larger approximant
unit cell contains more different local environments, reproduces larger
structural motifs from the quasicrystal, and should more closely approximate
the experimental sample.  Runs were performed on a sample of 8 unit cells
(2,744 atoms).  The cuboid-shaped tip was constructed of 340 FCC aluminum atoms
but treated as a rigid body.  In this case, to capture the compliance of an
AFM system, we connected the tip to a virtual atom using a harmonic spring
(spring constant 16 N/m) and moved the virtual atom across the substrate.
The sliding speed was 5 m/s and the target temperature 0K.  Figure
\ref{fig:keithT11} compares two different approximant surface terminations
to the atomic surface model derived from STM images in
Ref.~[\onlinecite{Park05b}].
On relaxation, both terminations yield Al-rich surfaces compatible with
the model; notable features include pentagons with a single vertex
exposed to the surface, distorted pentagons with two vertices on the
surface, and short and long motifs with lengths in the ratio of the
Golden Mean, as in the Fibonacci sequence.
The friction results are shown in
Fig. \ref{fig:keith1}.  Again, neither termination exhibits the experimental
anisotropy with larger friction in the periodic direction.  Other runs,
with sliding speeds as low 0.01 m/s, larger simulations (50,993 atoms),
different spring constants (16 or 8 N/m), and temperatures (0K or 300K)
also failed to reproduce the experimental anisotropy.\cite{McLaughlin09}

\section{Fibonaccium as a Generically Quasiperiodic Substrate}
\begin{figure} \begin{center}
      \subfigure[]{\includegraphics[scale=0.25]{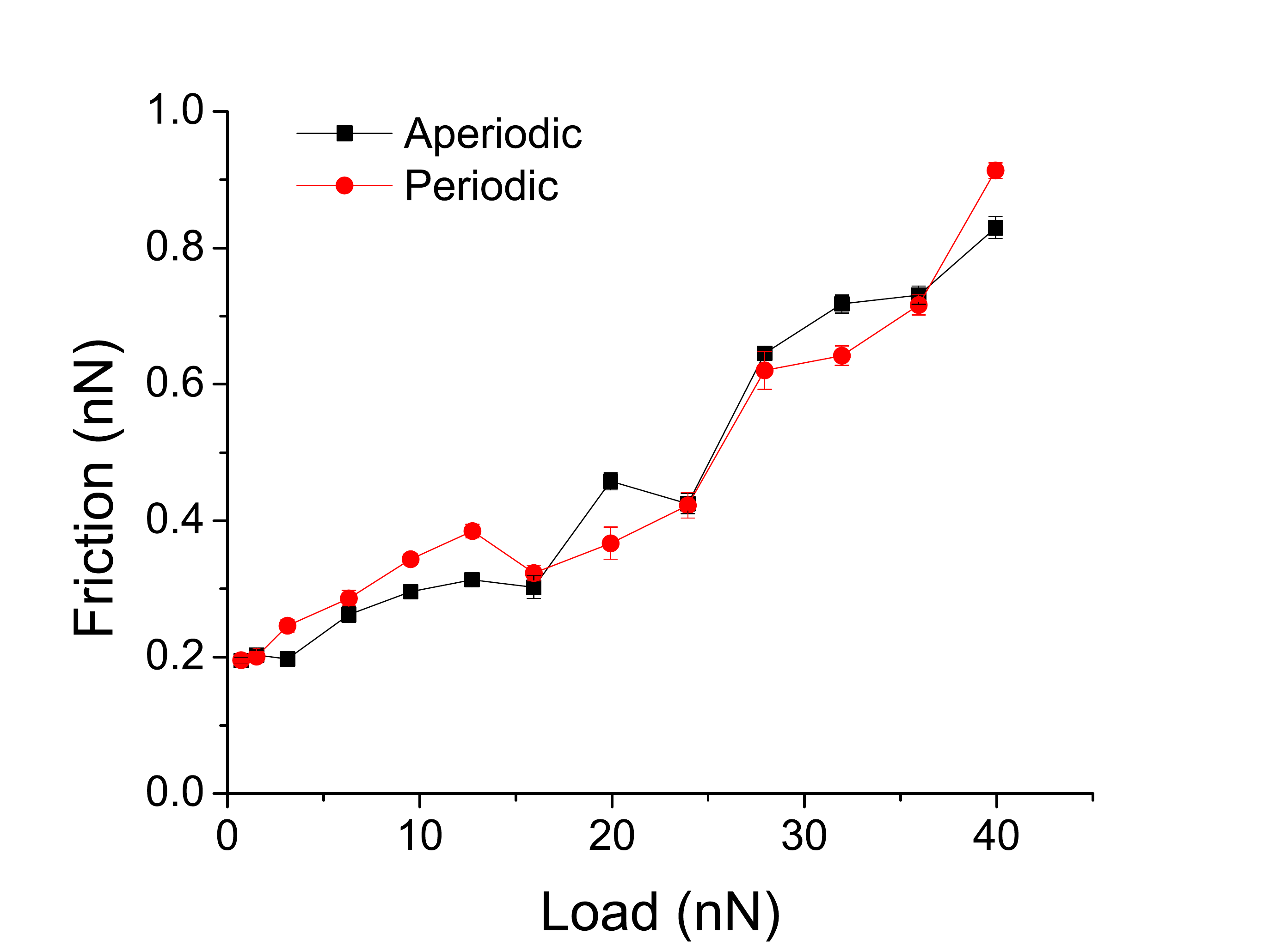}}
      \subfigure[]{\includegraphics[scale=0.25]{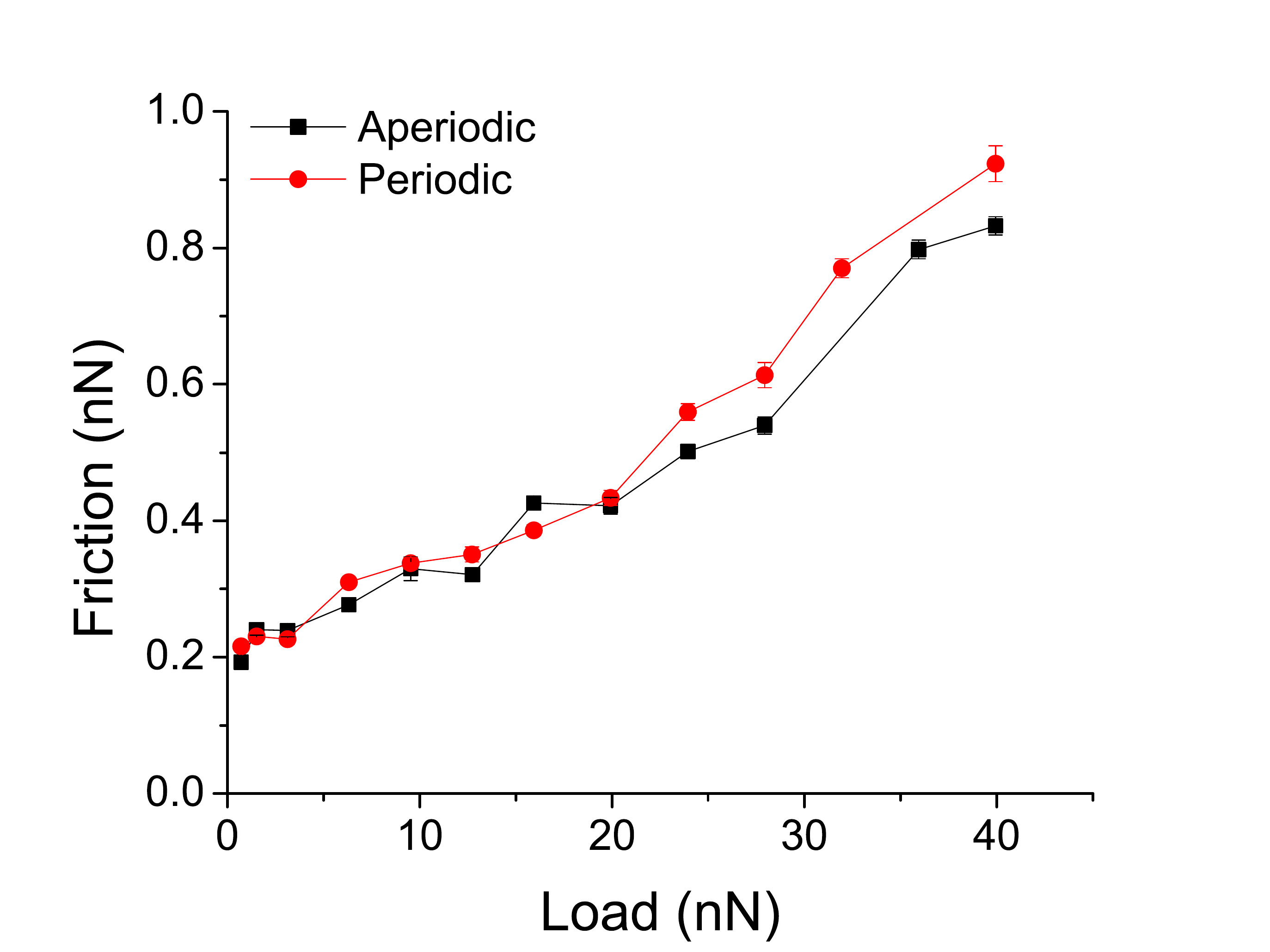}}
\end{center} \caption{Average friction force increasing with load from a
simulation of sliding on a substrate generated based on (a) period-3 and (b)
period-55 Fibonacci sequences. The frictional anisotropy of the experiments is
not captured by the simulation.} \label{fig:keith2} \end{figure}

Only a few realistic approximant structures
are available along with pair potentials that stabilize them, making it
difficult to increase unit-cell size systematically in the approach to
quasperiodicity.  One way around this difficulty is to simulate an
entirely artificial substrate for which approximants are easily generated.
The Fibonacci sequence
has been used in various studies as a model of quasiperiodicity
\cite{Boettger90,engel2007fibonacci,li2008fibonacci,pattnaik1992fibonacci}
and in some cases has been expanded to the two-dimensional case.
\cite{ilan2004fibonacci,urban2004fibonacci}
Adopting this geometry, we construct a ``Fibonaccium'' solid in which the atoms
occupy the sites of a simple-cubic lattice
but vary in mass according to the Fibonacci sequence (...LSLLS...) in
two dimensions, resulting in a large mass ($m_{LL}$) on grid sites
for which both coordinates
are $L$, a small mass ($m_{SS}$) where both coordinates are $S$,
and an intermediate mass ($m_{LS}=m_{SL}=[m_{LL}+m_{SS}]/2$)
on grid sites with $L$ in one coordinate
and $S$ in the other.  These two-dimensional grids stack periodically in the
third direction.
The artificial structure is stabilized with a simple pair potential.
Details of the approach are available elsewhere.\cite{McLaughlin09}
We varied the order of the Fibonacci approximant, temperature, sliding speed,
and spring constant, but no combination of these parameters yielded the
experimentally-observed anisotropic friction.  Representative results from simulations for
samples generated using period-3 and period-55 Fibonacci sequences are shown
in Fig. \ref{fig:keith2}.  If this were a real layered material, the linear
period of 55, corresponding to a two-dimensional unit cell of 3,025,
would be quite large and so substitutes for exact quasiperiodicty.
However, the experimental anisotropy is not present.

\section{Bare Al Tip on H1 Approximant}
\begin{figure} \begin{center}
      \includegraphics[scale=0.35]{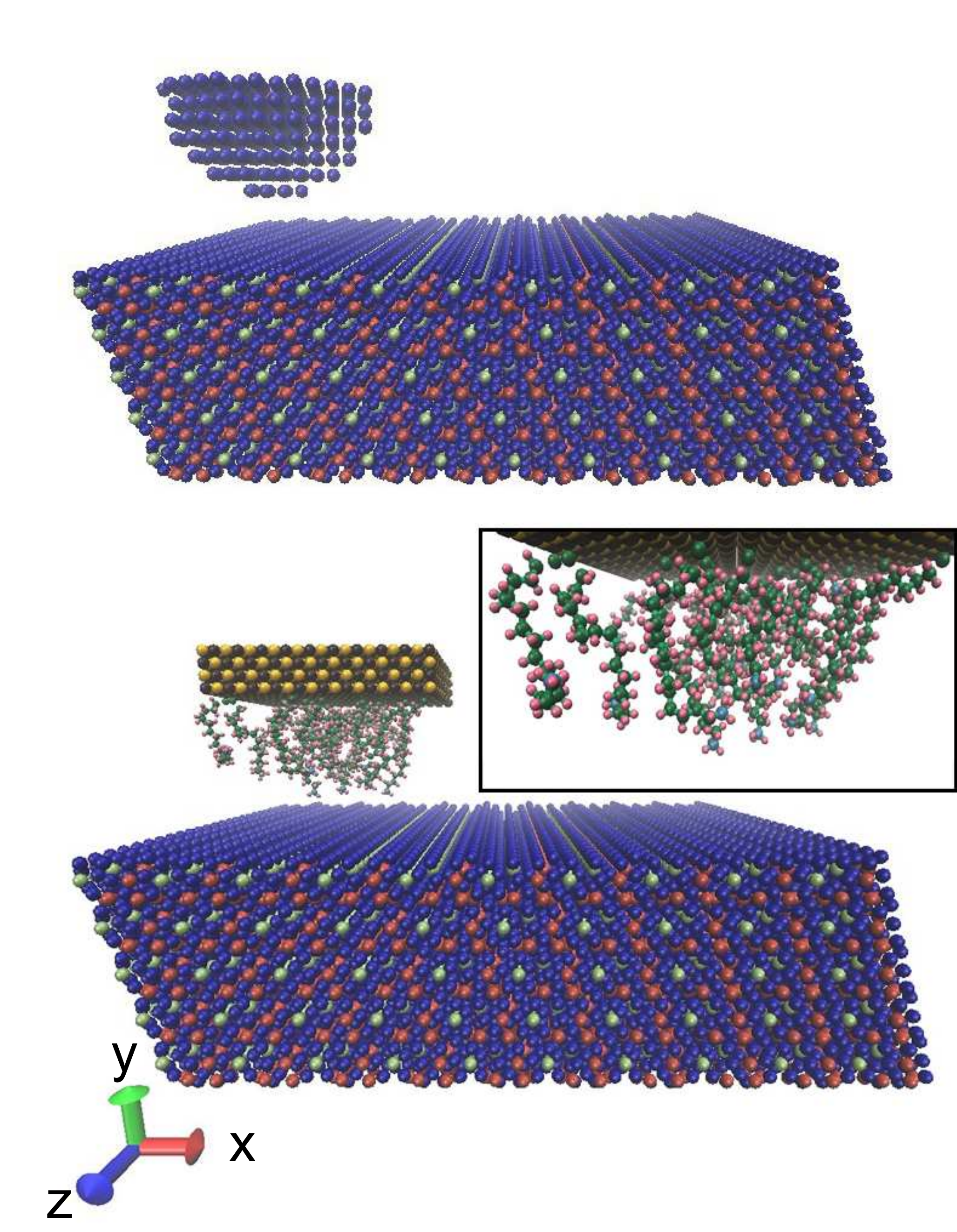}
\end{center} \caption{MD models of a hemispherical tip (upper image) and a
thiol-passivated TiN tip (lower image and inset) sliding on d-AlNiCo
substrate. Spheres represent atom positions where colors correspond to
different elements: yellow - Ti, black - N, dark green - C, light blue - S, pink - H, red - Co, green - Ni, blue - Al.} \label{fig:setup} \end{figure}

Seeing no anisotropy in the generic Fibonaccium model, we pursued instead
more realistic simulations of the experimental system.
New parameters to describe Al-Ni-Co alloys,\cite{mishin2015}
compounds, and mixtures using
the Embedded Atom Method (EAM) \cite{foiles1986embedded} were recently
made available via the
NIST Interatomic Potentials Repository Project \cite{NISTEAM}
based on the binary Ni-Al potential \cite{mishin2009NiAl} and elemental Co potential.\cite{mishin2012Co}
We used this model to describe the interactions within the previously described
H1-approximant sample.  The Al model tip was also modified by cutting atoms
from the original slab into a 2 nm radius hemisphere.  The uppermost atoms of
the tip were treated as a rigid body and pulled along the substrate at 5 m/s
by a virtual atom connected via a harmonic spring with spring constant 16 N/m.
This model is illustrated in Fig. \ref{fig:setup} (upper).
The bottom-most layers
of atoms in the substrate were fixed.  All unconstrained atoms in the system
were coupled to a Langevin thermostat in the two directions orthogonal to
sliding with a target temperature of 300K.

\begin{figure} \begin{center}
      \includegraphics[scale=0.35]{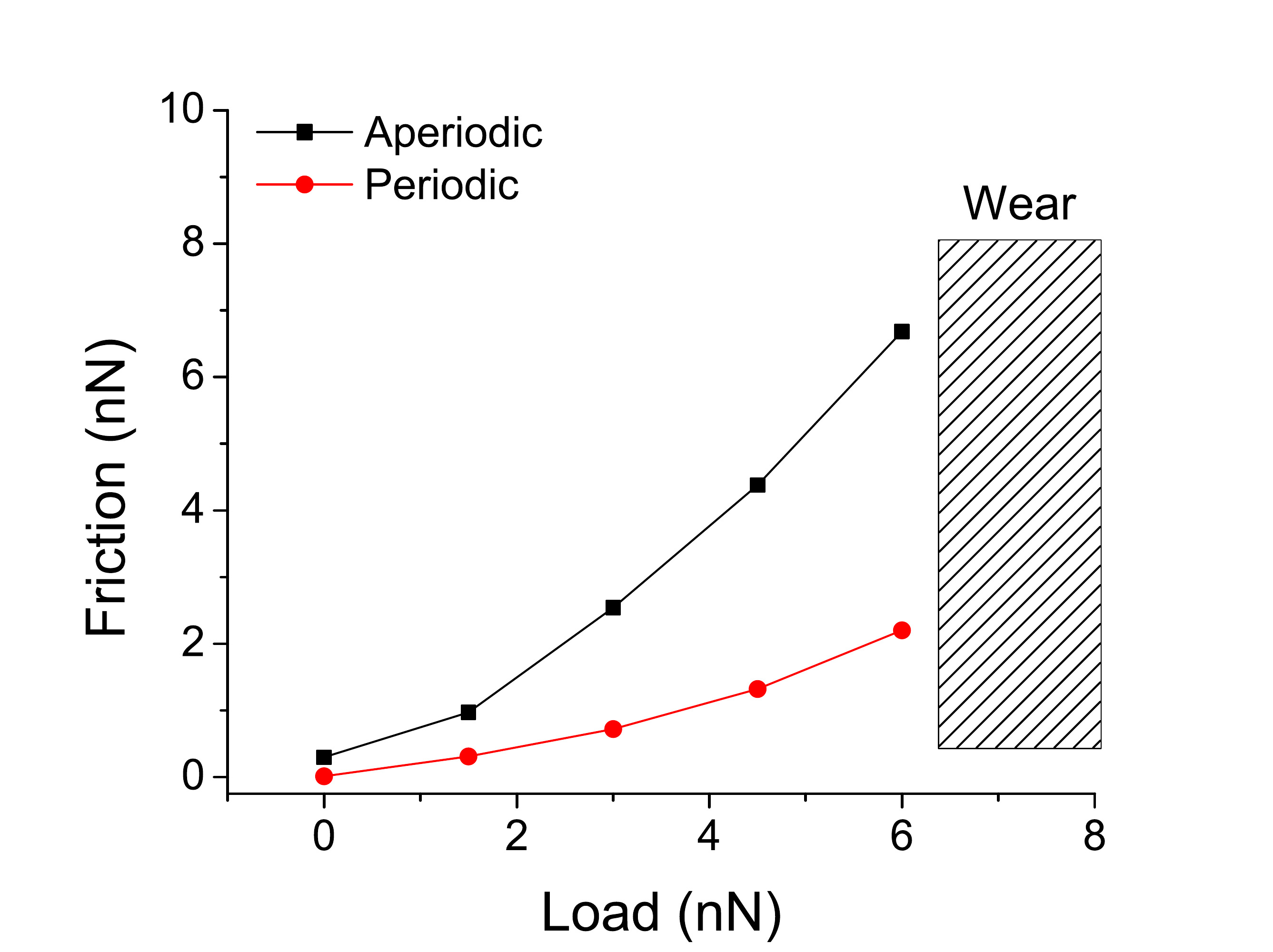}
\end{center} \caption{Average friction force increasing with load for a
hemispherical aluminum tip sliding on an H1 approximant substrate. Friction
magnitude is reasonable, but the frictional anisotropy is opposite to that
observed in previous experiments. Wear was observed at even moderate loads,
precluding characterization of friction above $\sim$ 6 nN.} \label{fig:justin1}
\end{figure}

Fig. \ref{fig:justin1} shows the average friction of the hemispherical tip
sliding on the H1-approximant substrate at different loads.  We observed
that, while the simulation captures frictional anisotropy, it is the opposite
direction of that observed in experiments, {\it i.e.}, friction is smaller in
the periodic direction.  In addition, we observed significant substrate wear,
even at moderate loads, and so could not run the simulation at loads larger than
$\sim$6 nN.

\section{Thiol-Passivated Tip}
The observation of wear from a bare tip is actually consistent
with experimental observation and is also the reason that the
experiments used a TiN tip passivated with hexadecane thiol,
\cite{park2005high,park2006tribological,park2008friction} eliminating wear, as
verified by STM imaging before and after.  To capture this in the simulation,
we constructed a new model of a thiol-passivated TiN tip to mimic that used
in experiments.  The model system is illustrated in Fig.~\ref{fig:setup}(b).
We first created models of a block of TiN (40$\times$10$\times$40 \AA) and 20
thiol molecules using Accelrys Materials Studio.  Then, we transferred the
structures into LAMMPS and artificially increased the interaction strength
between the bottom surface atoms of the TiN tip and the carbon atom at one
end of each thiol molecule so that one end of the thiol molecules was attached
to the TiN bottom surface after equilibration.  The TiN slab was subsequently
treated as a rigid body.  The Polymer Consistent Force Field (PCFF) was used
to describe bond, angle, torsion, and out-of-plane interactions between all
tip (TiN and thiol) atoms, the EAM potential was again used to describe the
inter-atomic interactions between substrate atoms, and the Lennard-Jones
(LJ) potential was used to model the long-range interactions between tip and
substrate with parameters obtained using Lorentz-Berthelot mixing rules.  All other
simulation parameters and conditions were the same as described for the
hemispherical Al tip model.

\begin{figure} \begin{center}
      \includegraphics[scale=0.35]{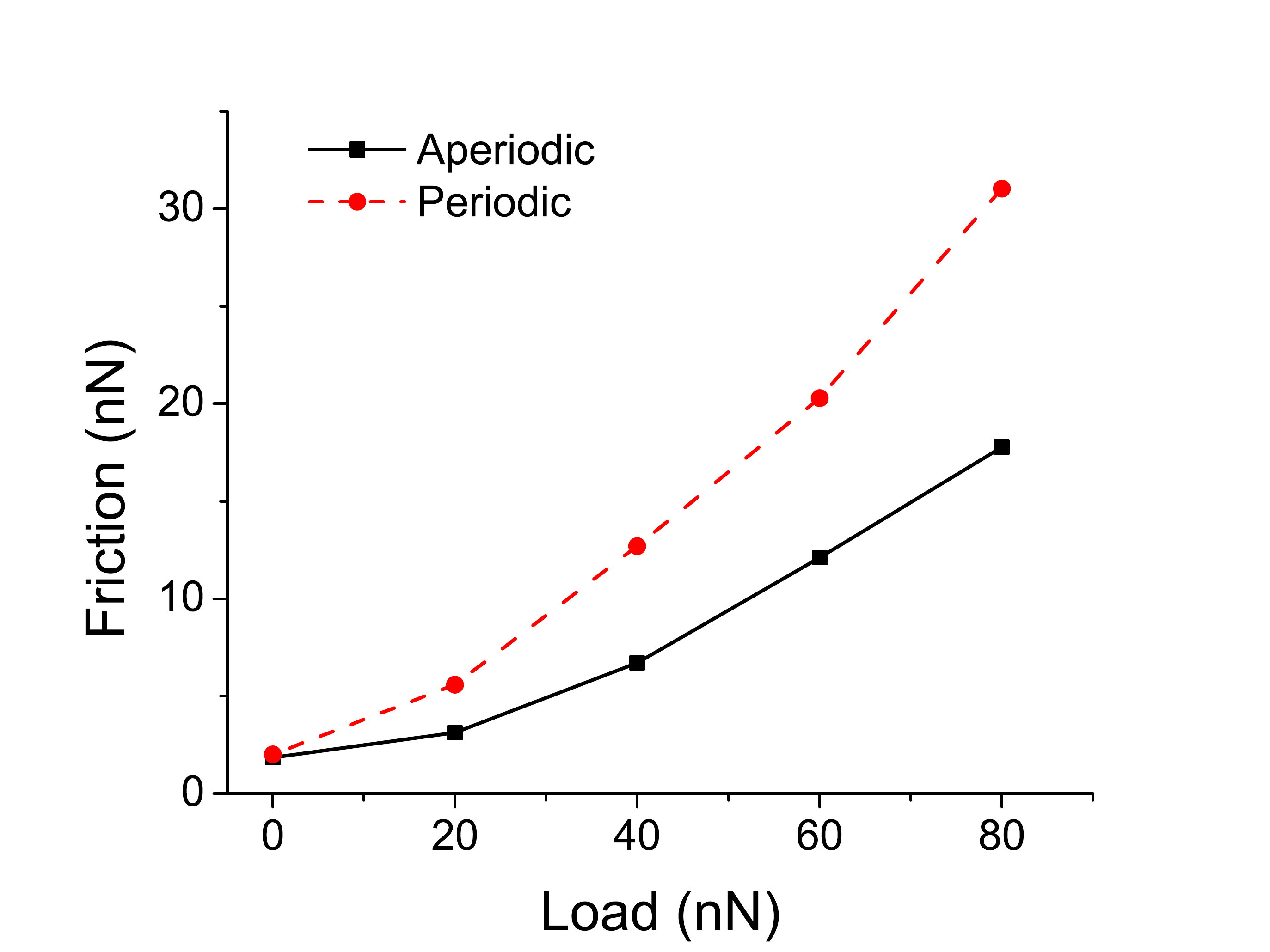}
\end{center} \caption{Average friction force increasing with load for a
thiol-passivated TiN tip sliding on the H1 approximant substrate.
No surface wear
is observed, and both the magnitude and the anisotropy of the
friction are roughly
consistent with experiment.}
\label{fig:lowloads} \end{figure}

Fig.~\ref{fig:lowloads} shows the average friction of the thiol-passivated TiN
tip sliding on the H1-approximant substrate at different loads.  We observed
higher friction in the periodic direction than the quasiperiodic direction,
consistent with the trend observed in experiments.  The magnitude of
the friction and friction coefficient were also roughly comparable to
the experiments, although the anisotropy was much smaller.
In this case, we did not observe any noticeable wear on the surface,
confirming that the thiol molecules indeed contributed to reducing wear.

\begin{figure} \begin{center}
      \includegraphics[scale=0.35]{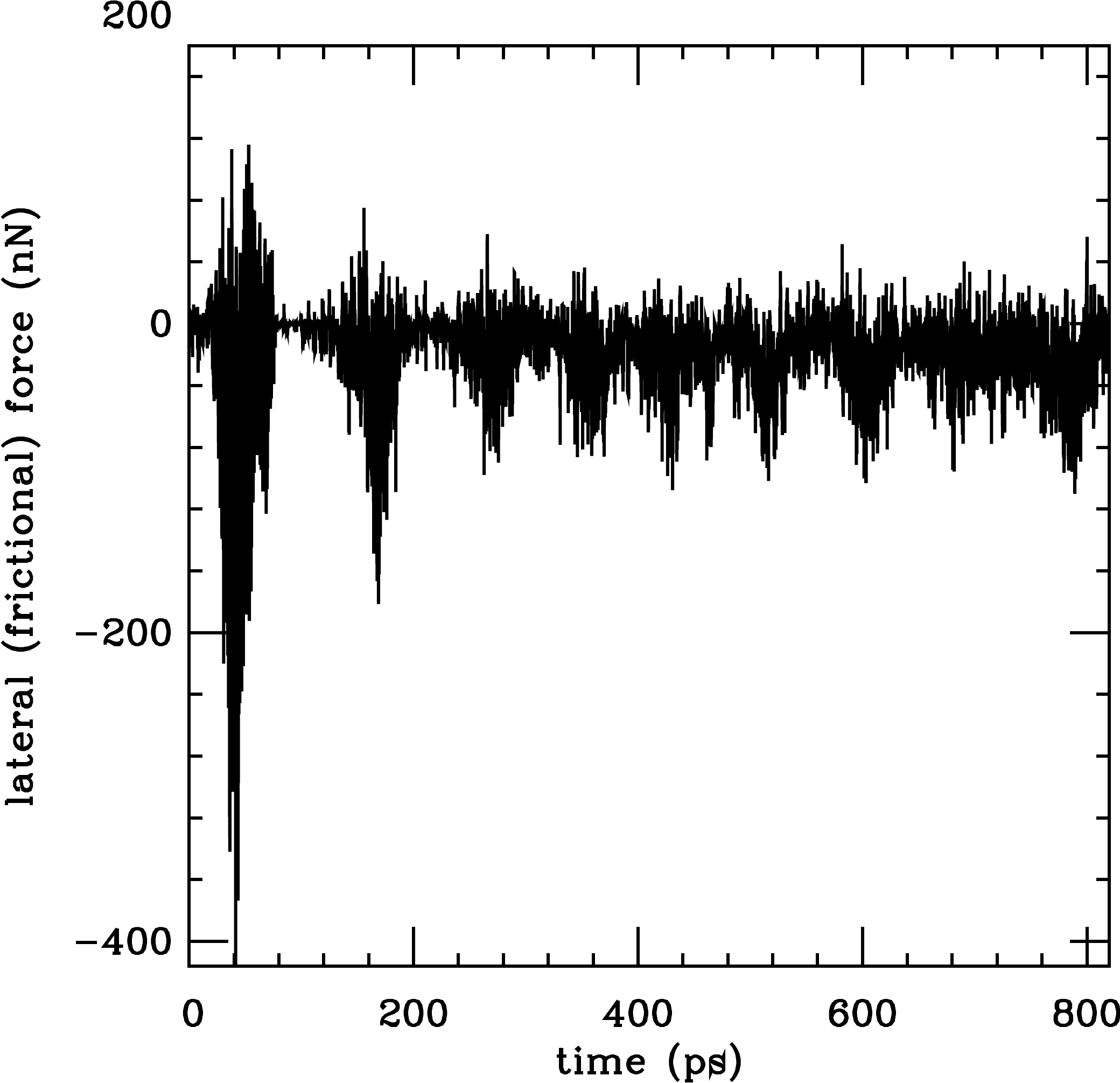}
\end{center} \caption{Sample run from Figure~\ref{fig:lowloads} for
the quasiperiodic direction at 80 nN load.  The time series is
noisy, but the pseudoperiod of $\sim82$--$98$ ps enables us to estimate error bars of
smaller than 1.5 nN to the average negative lateral force of 18.7 nN.  (The
first 400 ps are excluded in computing the average.)}
\label{fig:aperiodic-non-rigid} \end{figure}

Since frictional measurements are noisy, verifying
the anisotropy in Figure~\ref{fig:lowloads}
requires an analysis of the uncertainties.  Figure~\ref{fig:aperiodic-non-rigid}
shows a sample run at 80 nN load in
the quasiperiodic direction: the noise is large compared to the time-averaged
friction.  A pseudo-period of $\sim98$ ps, possibly reminiscent of
stick-slip friction, is evident in the time series; a similar
pseudo-periodicity was observed in the periodic direction.
Although experimentally
stick-slip has been reported absent or suppressed at low loads in
true quasicrystals,\cite{Mancinelli03,Park05c,park2006tribological}
we would
still expect it in this low-order approximant.  A power-spectral estimate
shows a broad peak closer to $82$ ps than $98$ ps, and we shall use this
periodicity in estimating error bars.  At the sliding speed of 5 m/s, this
periodicity also corresponds closely to the approximant unit-cell size of
$4.03$~\AA, although in other runs, there were differences between the
stick-slip-like pseudo-periodicity and corresponding unit-cell dimension
as large as 20\%.

Fig.~\ref{fig:aperiodic-non-rigid}
suggests initial transient behavior that subsides by 300 ps.
In these runs, we averaged only data past 400 ps.  Table~\ref{tab:friction}
presents the mean friction values with error bars.  To estimate error bars,
we collected the data past 400 ps into $N$
bins each of duration equal to the pseudo-period so that the mean friction
in each
bin is considered an independent measurement.  The error bars are
then estimated as a sample standard deviation
of the bins divided by $\sqrt{N}$.  Since the pseudo-periods are themselves
uncertain, we compared the means and estimated error bars for
pseudo-periods 20\% larger, 20\% smaller, twice as large, and half as large
as that extracted from the power spectrum.  The table displays as ``worst''
the estimate with the largest error bar.  In each case, the results
show a highly significant anisotropy of order 60\%--75\% with negligible
sensitivity to the binning procedure.

\begin{table}
\caption{Frictional forces under 80 nN load with estimated uncertainties.
Runs with rigid substrates eliminate lattice-vibration effects.
For each type of substrate (non-rigid and rigid), the first row gives
the frictional forces along the periodic and quasiperiodic directions.
Friction and error bars in nN are estimated from equal-duration bins
given by the pseudo-period, with the error bar the standard deviation of
the mean of the independent bins.
The row labeled ``worst'' for each substrate type substitutes the pseudo-period
and error bar derived from an alternative binning procedure in which the bins
are either as given previously, 20\% larger or smaller, or twice or half
as large, {\it whichever yields the largest error bar}.
For both substrates, the periodic friction is
larger than the quasiperiodic by at least seven standard deviations of the mean.
}
\setlength{\tabcolsep}{1em}
\begin{tabular}{lr@{.}l r@{.}l}%\\
substrate & \multicolumn{2}{c}{periodic friction}& \multicolumn{2}{c}{quasiperiodic friction}\\
\hline
non-rigid & 30 & 8(1.0) & 18 & 7(0.9)\\
{\it worst} & \it30 & \it8(1.7) & \it18 & \it4(1.5)\\
\hline
rigid & 39 & 2(1.5) & 22 & 1(1.2)\\
{\it worst} & \it39 & \it2(1.5) & \it22 & \it1(2.1)\\
\label{tab:friction}
\end{tabular}
\end{table}

One hypothesis for the origin of the frictional anisotropy has involved the
difficulty of exciting or propagating phonons through the quasicrystal in
the quasiperiodic direction.  We can test this by suppressing lattice vibrations entirely using simulations where the atoms in the substrate are artificially fixed in place after equilibration.
If phonons were
important, we might expect the anisotropy to decrease for the rigid substrate
and for the overall friction to be lower.
Table~\ref{tab:friction} shows exactly the opposite: the overall friction
increases 20--30\%, while the anisotropy remains approximately the same or
increases slightly on the rigid substrate.  With the substrate atoms fixed,
the only degrees of freedom left are in the thiol-passivated tip.
The difference in friction between the two cases could result
from the depression of surface asperities in the non-rigid case, something
we see when we plot atomic positions.  On both substrates, however,
the friction was higher in the periodic direction.

\begin{figure} \begin{center}
      \subfigure[]{\includegraphics[scale=0.25]{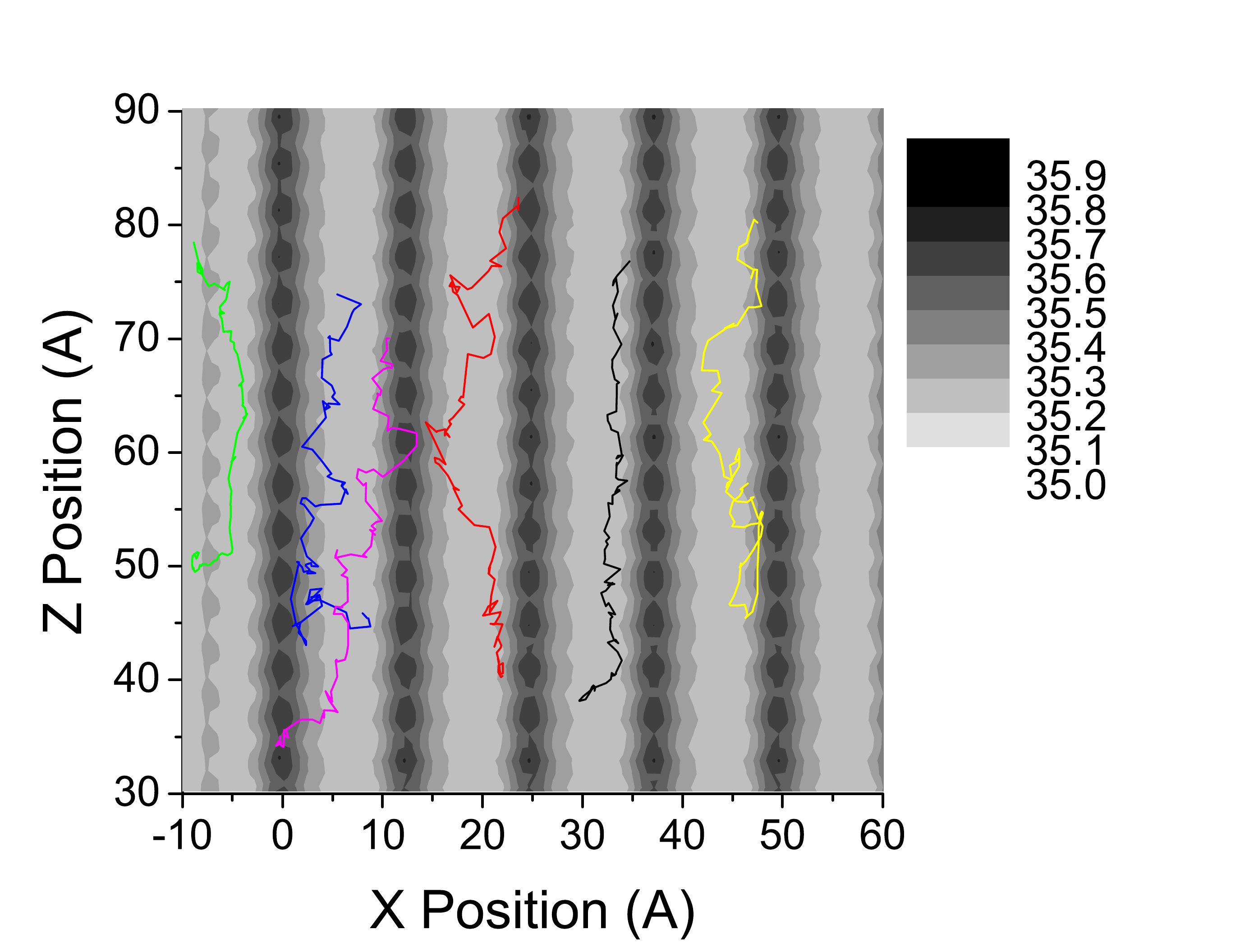}}
      \subfigure[]{\includegraphics[scale=0.25]{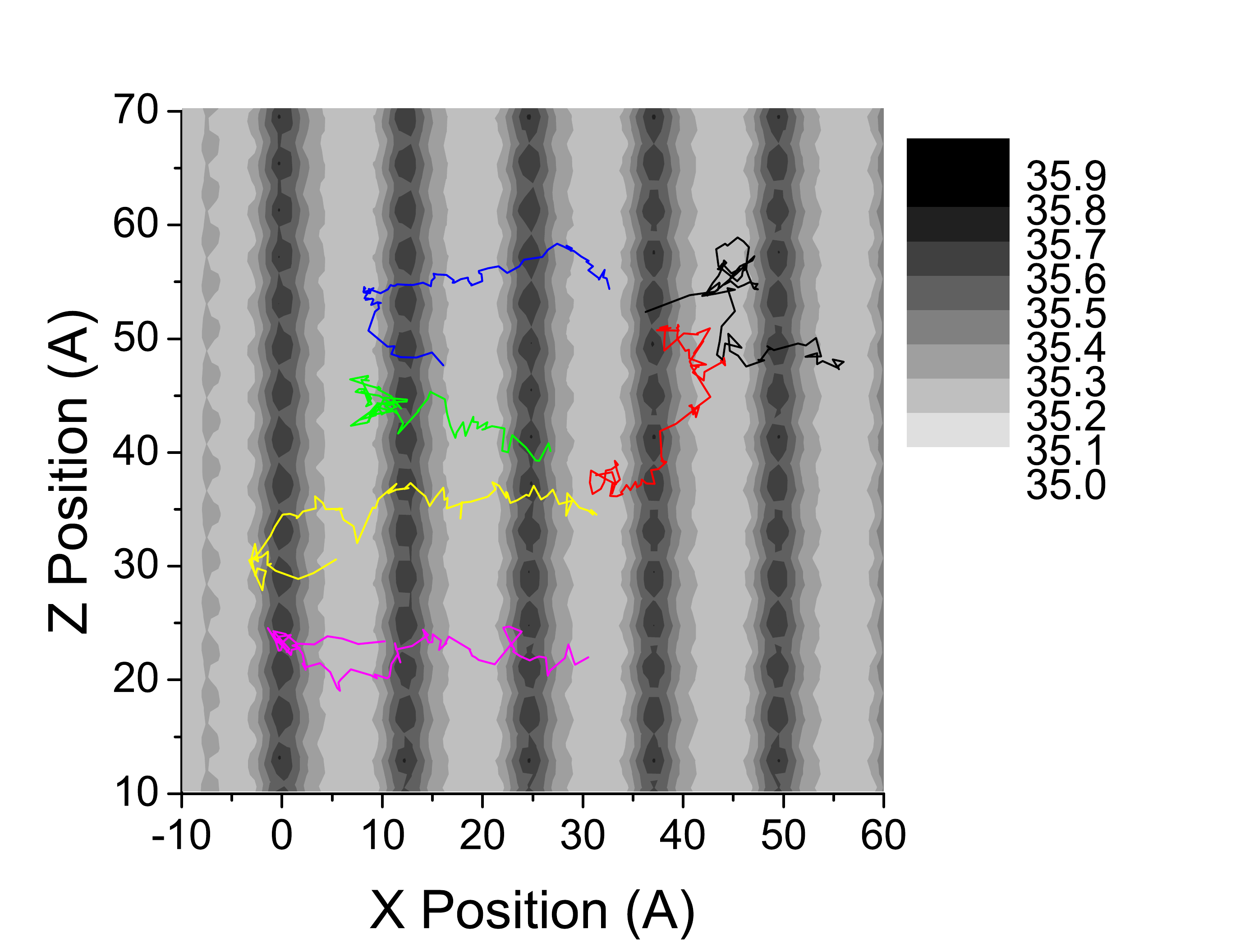}}
\end{center} \caption{The trajectories of the last carbon atom in the thiol
chains when the tip slid in the (a) $z$-direction (periodic) and (b) $x$-direction
(quasiperiodic).
The $z$-periodicity is 4.03\AA, while
the periodicity of the approximant in the quasiperiodic $x$ direction is
12.22\AA.
The gray-scale maps show the vertical heights of the quasicystal
surface in Angstroms. } \label{fig:f3} \end{figure}

In order to understand why the thiol-passivated tip model was successfully able
to capture the expected friction trends, we characterized the trajectory of
the last carbon atom in the thiol chains as the tip slid.  Figure \ref{fig:f3}
illustrates the vertical height distribution of the quasicystal surface as
gray-scale maps, where the lighter gray corresponds to a lower vertical
position.
We can identify parallel furrows (low vertical positions) and peaks (high
vertical positions) in the gray map due to the periodic surface structure
of the quasicrystal.  The colored lines indicate the trajectories of the last
carbon atom in six representative thiol chains.  We observed very different
trajectories during sliding in the periodic and quasiperiodic directions.
Most significantly, for sliding in the periodic direction (z-direction)
illustrated in Fig. \ref{fig:f3}(a), the trajectories remained, on average,
in the light colored regions on the gray-scale surface-height map, indicating
that the chains were entrained in the surface furrows.  This entrainment
could contribute to adhesive friction.
In contrast, for sliding in the quasiperiodic direction (x-direction)
illustrated
in Fig. \ref{fig:f3}(b), the thiol chain tail trajectories exhibited irregular
motion patterns corresponding to limited chain entrainment.

\begin{figure}
\begin{center}
\subfigure[]{\includegraphics[scale=0.39]{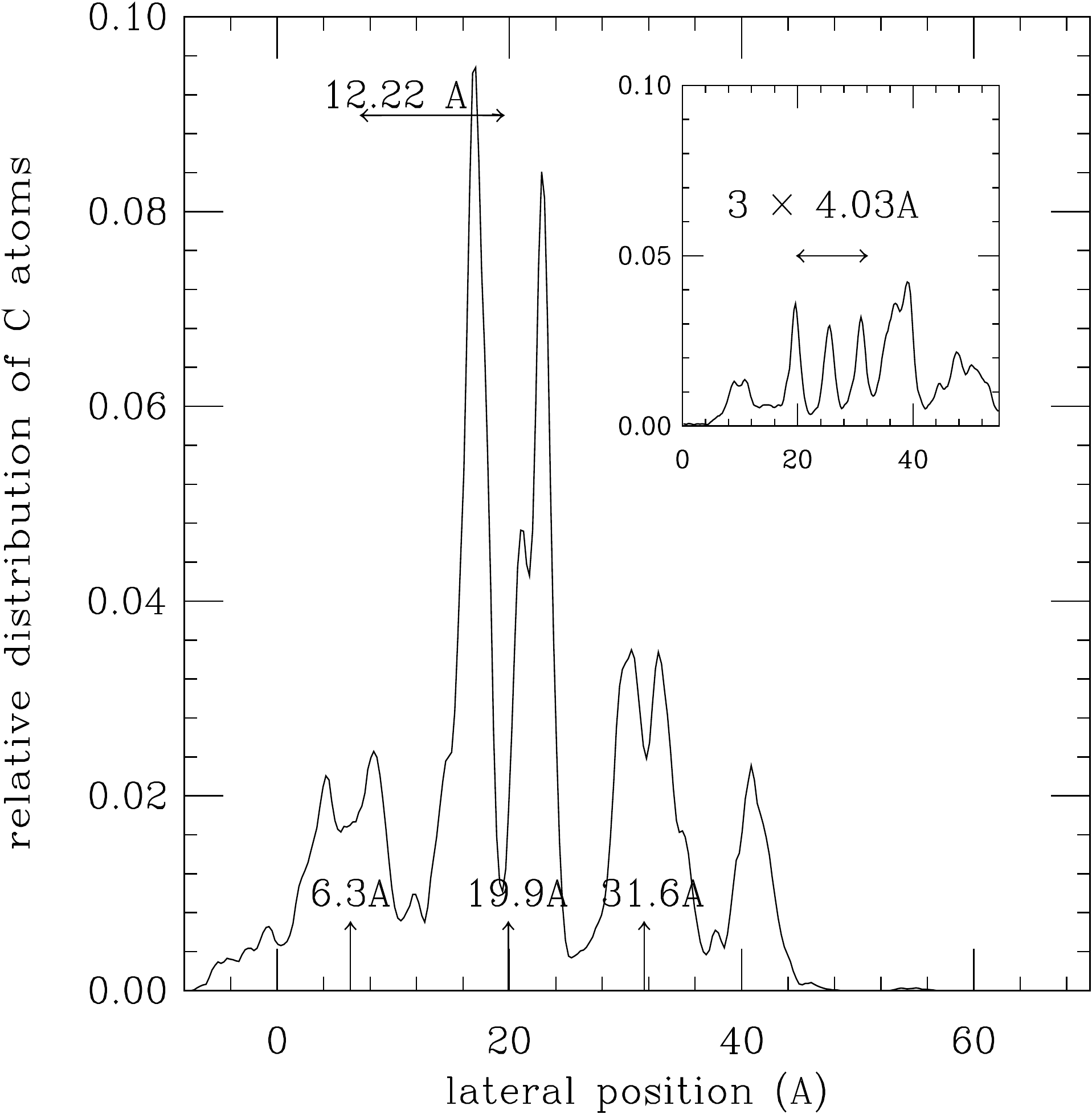}}
%\subfigure[]{\raise1.40\baselineskip\hbox{\includegraphics[angle=-90,origin=c,scale=0.31]{altcrop.pdf}}}
\subfigure[]{\raise1.40\baselineskip\hbox{\includegraphics[scale=0.31]{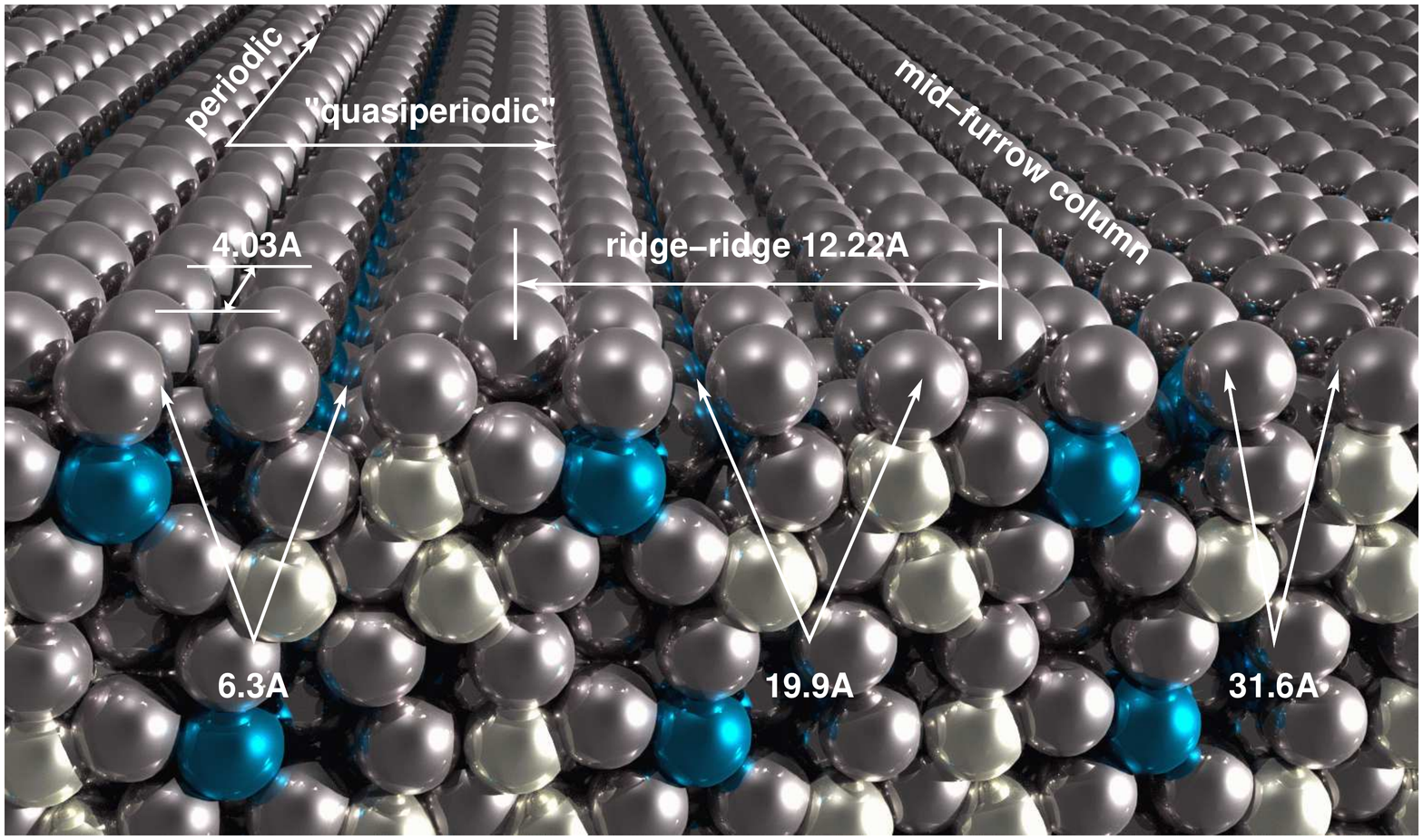}}}
\end{center}
\caption{\baselineskip0.75\baselineskip (a) Relative distribution of lateral position of carbon
atoms during the second
half of the simulation dragging in the periodic direction ({\it i.e.}, the
position axis is along the {\it quasiperiodic\/} direction.)  Only
atoms with absolute height less than 38~\AA, about 2.6~\AA ~above
the nominal surface height, are counted.  Split peaks centered at
6.3~\AA, 19.9~\AA, and 31.6~\AA\ are consistent
with carbons localized near furrow edges, where the furrows are spaced
12.22~\AA\ apart.
The smoothing kernel has a standard
deviation of 0.3~\AA.  Inset: distribution of lateral position for
an aperiodic run (so the position axis runs along the periodic direction).
Several pairs of adjacent peaks are spaced approximately 5.7 \AA\ apart, not
corresponding to any surface features or to the 4.03 \AA\ periodicity.
(b) Quasicrystal approximant surface, showing approximant periodicities
of 12.22 \AA\ and 4.03 \AA\
and furrows
aligned along the periodic direction (into the picture, periodicity 4.03 \AA).
Each furrow is divided in two by a column of Al atoms, while the walls of the
furrows consist of a higher ``ridge'' of three more tightly packed
Al columns.
Arrows mark the split peaks in Fig.~\ref{fig:furrowC}a, labeled by the
position half-way between the two parts and
corresponding to the centers of shallow furrows as illustrated here and
in the height maps of Fig.~\ref{fig:f3}.
Note five-fold motifs on the ten-fold surface, facing forward.
\label{fig:furrowC}} \end{figure}

Fig.~\ref{fig:furrowC} provides additional evidence for entrainment of
thiol chains in furrows when the tip is dragged in the periodic direction.
This shows a kernel-smoothed distribution (kernel width 0.3$\,$\AA) of
lateral carbon positions for all carbon atoms
at heights less than about 2.6$\,$\AA ~above the highest nominal (undepressed)
surface-atom center,
excluding the first 400 ps
of the simulation.  Split peaks centered at 6.3$\,$\AA, 19.9\,\AA,
and 31.6$\,$\AA ~are roughly consistent with the known furrow centers at
lateral positions $(6.61+12.22n)$\AA\ for integer $n$, where 12.22$\,$\AA\ %
is the approximant periodicity.  Figures \ref{fig:f3} and \ref{fig:furrowC}b
show relatively shallow furrows with centers near these positions.  A
column of Al atoms divides each furrow in two.
The splitting of the
peaks could be due to this column or indicate affinity of carbons
for the furrow sides.
%, although
%the widths are not quite equal.
%A similar distribution for a tip
%dragged in the quasiperiodic direction shows
%periodic but less pronounced peaks.
We count
carbon atoms at heights less than or equal to the nominal surface height
as ``in'' the furrows and find 3.6 times as many such carbons when the
tip is dragged in the periodic direction as when dragged in the quasiperiodic
direction.  
%The inset to Fig.~\ref{fig:furrowC} plots the distribution of the smaller
%number of carbons below the nominal surface during a run in the quasiperiodic
%direction.  While peaks still appear, their spacing does not correspond
%to multiples of the 4.03$\,$\AA periodicity in the periodic direction.
%(Several paris of peaks are separated by about 5.6$\,$\AA; since this

\begin{figure} \begin{center}
      \subfigure[]{\includegraphics[scale=0.28]{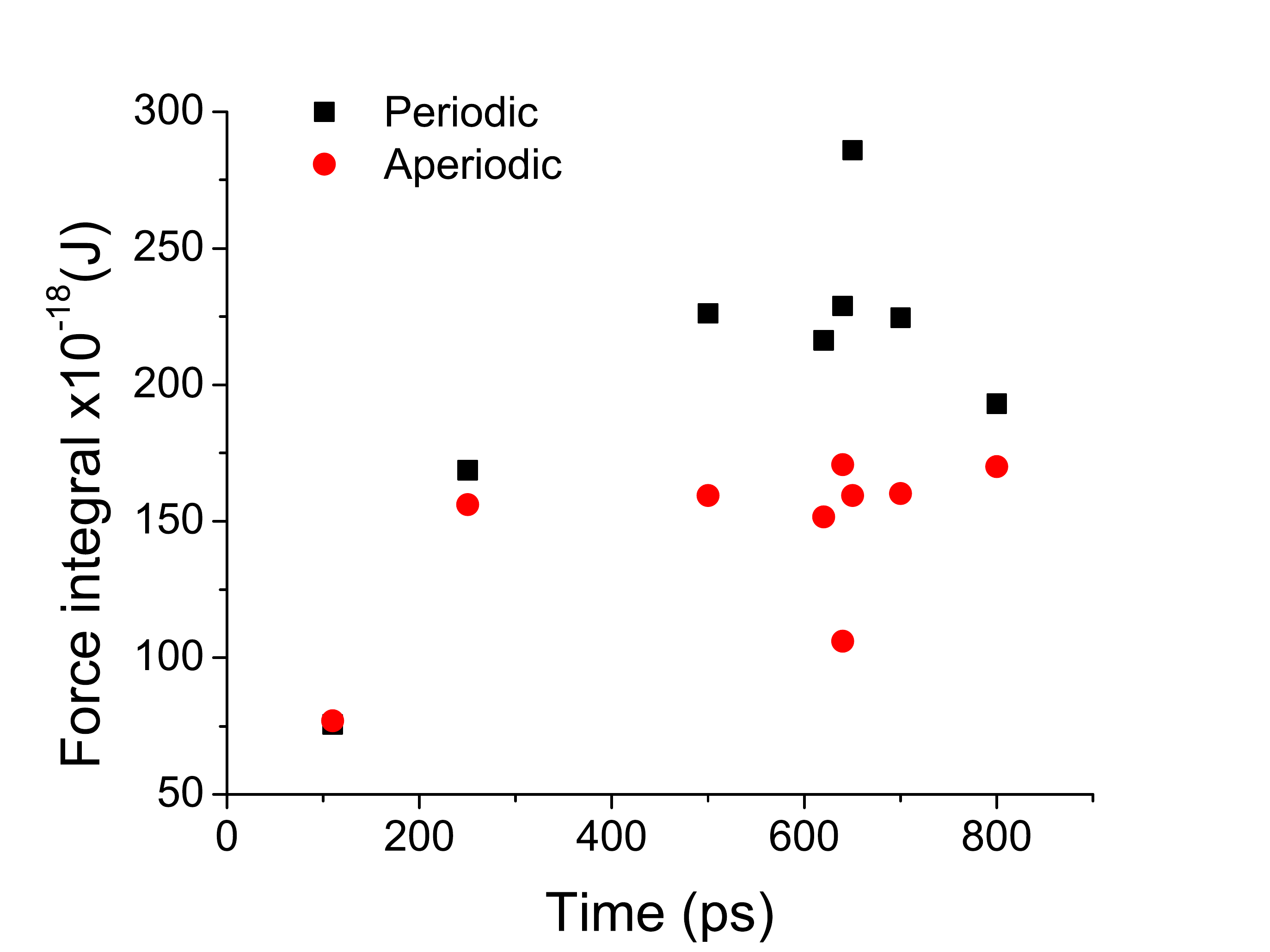}}
      \subfigure[]{\includegraphics[scale=0.28]{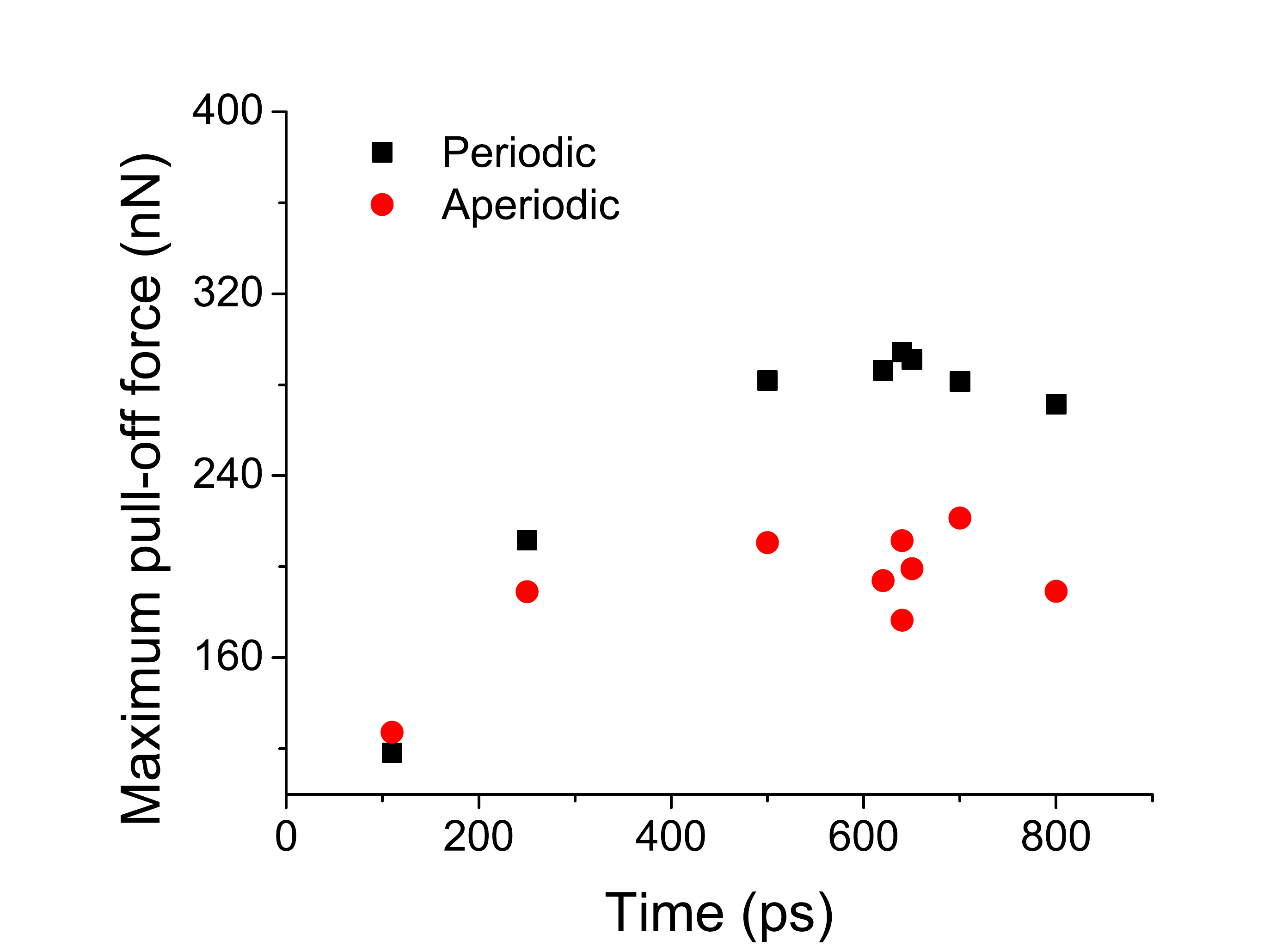}}
\end{center} \caption{Adhesive energies (left) and maximum adhesive forces
during lift-off (right) after runs at 80 nN load of durations
between 100~ps and 800~ps.  Since the aluminum tip is starting from the
same height in all cases, any differences must pertain to the thiol chains.
The graphs show greater
adhesion for runs in the periodic direction, consistent with entrainment of
the chains in furrows.
} \label{fig:adhesive} \end{figure}

Figure~\ref{fig:adhesive} shows adhesive energies and
maximum forces for runs at 80 nN load stopped at times between
100~ps and 800~ps.  After the sliding is stopped, the tip is separated
from the substrate, up to a height of 3.2~nm (somewhat longer than the length of
a thiol chain, 2.4~nm).  We calculated the force on the tip as a function of distance from the substrate during lift-off. The left half of the figure shows total integrated
adhesive energy, the right half, the maximum force.
Both show substantially greater adhesion after runs in the periodic direction.
%consistent with the idea of entrainment in
%furrows.
Both also level off by 400 ps, supporting our previously cited
evidence that a steady state is reached by that time.
Since the aluminum tip starts off at the same height above the surface in
all cases, the differences in measures of adhesion must pertain to the
thiol chains.  We note that after short runs, there is essentially no
difference in lift-off force or integrated adhesive energy, but after the
chains have stretched out along the surface and, as we have argued,
find themselves entrained in the furrows, the anisotropy becomes clear.
Entrainment of thiols in furrows and the consequent higher adhesive force
when the tip is dragged in the periodic direction could contribute to
increased friction.
%Experimental measurements of adhesive force
%\cite{Park04,Park05c,park2006adhesion} on the
%two-fold and ten-fold surfaces of d-AlNiCo were performed by lowering
%and raising of the tip in the normal direction rather than after dragging
%the tip across a surface, so a direct comparison to experiment is not possible.

\section{Discussion}

Explaining the giant frictional anisotropy observed on the two-fold surface
of d-AlNiCo requires attention to realistic details; generic models
showed no anisotropy, or showed reversed anisotropy, or proved
sensitive to parameters.
Comparison of the two terminations of the T11 approximant,
in which the less stable termination appeared to show a substantial reversed
frictional anisotropy (quasiperiodic direction with higher friction) demonstrates
%%%the importance of coordinating with energetic considerations.
an unanticipated dependence on details of surface structure.

Moreover, in order to avoid surface damage, it is necessary to minimize adhesive forces,
either artificially through the use of purely repulsive pair potentials (as in
``adamant'') or directly by simulating the thiol passivation.  We have seen
that the former results in unrealistically small frictional forces and may
miss important physics.

Entrainment of organic chains in surface furrows provides an intuitively
attractive
mechanism for anisotropy on this surface.  When the tip moves in the periodic
direction, parallel to the furrows, the chains preferentially stay aligned,
and their adhesion to the furrow sides may contribute to friction.
Conversely, when the tip moves
perpendicularly, the chains splay in apparently random configurations;
additionally, these configurations may change frequently as the chains move
across successive ridges in the (approximately) quasiperiodic direction,
minimizing adhesion.  Of course, this mechanism does not actually depend
on quasiperiodicity; the H1 unit cell realizes only the 1/1 approximant
to the golden mean.  The idea that local topographic features, rather
than quasiperiodicity, could
control the anisotropy has also been advanced by Filippov {\it et al.}.
\cite{filippov2010origin,filippov10E,filippovreply11}
Using a Langevin model over an external potential
representing the substrate, rather than molecular dynamics, the authors
suggest that larger force gradients on the atomic scale in the periodic
direction could explain the anisotropy.  However, the parameters in that
model have been questioned,\cite{mclaughlin11} and small changes can lead
to a reversed anisotropy.  The furrow model on the other hand, makes no {\it
ad-hoc\/} assumptions about effective atomic shapes and should be robust
against uncertainties in structural determination.

We note a possible connection between approximant quasiperiodicity and
the existence of surface furrows that will survive the limit of perfect
quasiperiodicity.  Spacings between rows, as in the surface Al atoms
in Fig.~\ref{fig:furrowC}b, follow the Fibonacci sequence (...LSLLS...),
as suggested even in the approximant of the figure.  If the short (``S'')
spacings tend to squeeze rows and push them up relative to the long, we
could arrive generically at the sort of surface corrugation invoked here.

The furrow model does not
come close to the eight-fold observed anisotropy in
frictional force.  Sliding velocities in these simulations were much larger
than in the experiment, and it is possible that one would see larger effects
at realistic speeds, if they could be achieved.

However, any model relying solely on local topographic features fails to
answer the question of why surface friction appears to be lower on the
doubly quasiperiodic surfaces of true quasicrystals than on approximants, and
lower on the approximants than on other phases in these alloy systems.  Nor
does it address
why quasicrystals show lower friction in engineering-scale pin-on-disk
experiments in air, with ploughing through an oxide layer, as well as in
nanoscale experiments in vacuum.\cite{park08b}

The Fibonaccium model was designed to test the idea that the difficulty
of either exciting or propagating phonons in a quasiperiodic direction
of a quasicrystal having both periodic and quasiperiodic directions could
account for the anisotropy.  These runs found no such effect, and although
theoretically phonon spectra in a quasicrystal should exhibit a dense set
of gaps,\cite{Boettger90,Hafner96} experiments find mostly isotropic
elastic properties.\cite{chernikov98PRL,Dugain99}
That friction, likewise, appears isotropic in a model capturing the
quasiperiodicity of real quasicrystals while ignoring their topography,
weakens the case for the phonon hypothesis.

There are at least three ways out.  First, none of the simulations attempted
to date considers electronic excitations, which may play a role.\cite{krim95}
Second, unrealistic aspects of molecular dynamics, such as scan speeds or
finite-size effects, may obscure physical mechanisms.

Finally, it may be that we're looking at two different effects.  The giant
frictional anisotropy on the two-fold surface of d-AlNiCo may in fact be due
entirely to local topographic features, such as thiol-chain entrainment, having
nothing to do with quasiperiodicity, while the relatively low friction of
quasicrystal surfaces compared to those of approximants, and of approximants
compared to other periodic phases, could be due to the greater hardness of
the quasicrystals and to a pseudogap in the density of
electronic states.\cite{dong99,rabson2012review}
Both are definite effects of quasiperiodicity for the quasicrystals and
of unit-cell size for the approximants.  Both are neatly factored out on
the two-fold quasicrystal surface, possibly leaving topography as the only
feature different in the two directions.

\section{Acknowledgments} 
We thank Susan Sinnott and Jacob Israelachvili
for useful suggestions and Dr.\ Gustavo Brunetto for generating
the ray-tracing image shown in Fig.~\ref{fig:furrowC}b.
Computations were performed on the CIRCE cluster
at the University of South Florida and the Coates/Conte/Carter clusters
at Purdue University.  The initial atomic configuration of the thiol tip
model was created with the assistance of U.S.\ Ramasamy.
AM and ZY acknowledge the support of the National
Science Foundation through grant CMMI-1362565.  PT acknowledges
support from the Office of Science, Basic Energy Sciences, Materials
Sciences and Engineering Division of the U.S.\ Department of Energy
(USDOE), under Contract No. DE-AC02-07CH11358 with the U.S.\ Department
of Energy.  DAR acknowledges NSF MRI award CHE 1531590 and the support of
the Research-Computing group at the University of South Florida.

\bibliography{reference}

\begin{thebibliography}{54}
\bibitem [{end()}]{endnote}%
  \BibitemOpen
  \href@noop {} {}\bibinfo {note} {An isotropic thermostat at an earlier stage
  of the investigation introduced significant noise; restricting the thermostat
  to the $z$ direction\cite{SinnottPrivate,Gosvami11} reduced the noise. If the
  phonon hypothesis had been correct, the lateral thermostat could have masked
  the effect.}\BibitemShut {Stop}%
\end{thebibliography}

\end{document}